\documentclass[aoas]{imsart}

\RequirePackage{amsthm,amsmath,amsfonts,amssymb,algorithm,algpseudocode,array,bm,bbm, booktabs,multirow,mathtools,soul,xcolor}
\RequirePackage[authoryear]{natbib}
\RequirePackage[colorlinks,citecolor=blue,urlcolor=blue]{hyperref}
\RequirePackage{graphicx}
\newcommand{\vect}[1]{\boldsymbol{\mathbf{#1}}}

\startlocaldefs

\endlocaldefs
\begin{document}

\begin{frontmatter}
\title{Bayesian spatial functional data clustering: applications in disease surveillance}

\begin{aug}
\author[A]{\fnms{Ruiman }~\snm{Zhong}\ead[label=e1]{ruiman.zhong@kaust.edu.sa}\orcid{0000-0000-0000-0000}} ,
\author[A]{\fnms{Erick}~\snm{A. Chacón-Montalván}\ead[label=e2]{erick.chaconmontalvan@kaust.edu.sa}}
\and
\author[A]{\fnms{Paula}~\snm{Moraga}\ead[label=e3]{paula.moraga@kaust.edu.sa}}
\address[A]{Computer, Electrical and Mathematical Science and Engineering Division, King Abdullah University of Science and Technology (KAUST), Thuwal, 23955-6900, Makkah, Saudi Arabia \printead[presep={,\ }]{e1,e2,e3}}
\end{aug}

\begin{abstract}
The ability to accurately cluster contiguous regions with similar disease risk evolution is crucial for effective public health response and resource allocation. In this article, we propose a novel spatial functional clustering model designed for disease risk mapping, utilizing random spanning trees for partitioning and latent Gaussian models for capturing within-cluster structure. This approach enables the identification of spatially contiguous clusters with similar latent functions, representing diverse processes such as trends, seasonality, smooth patterns, and autoregressive behaviors. Our method extends the application of random spanning trees to cases where the response variable belongs to the exponential family, making it suitable for a wide range of real-world scenarios, including non-Gaussian likelihoods.
The proposed model addresses the limitations of previous spatial clustering methods by allowing all within-cluster model parameters to be cluster-specific, thus offering greater flexibility. Additionally, we propose a Bayesian inference algorithm that overcomes the computational challenges associated with the reversible jump Markov chain Monte Carlo (RJ-MCMC) algorithm by employing composition sampling and the integrated nested Laplace approximation (INLA) to compute the marginal distribution necessary for the acceptance probability. This enhancement improves the mixing and feasibility of Bayesian inference for complex models.
We demonstrate the effectiveness of our approach through simulation studies and apply it to real-world disease mapping applications: COVID-19 in the United States of America, and dengue fever in the states of Minas Gerais and São Paulo, Brazil. Our results highlight the model's capability to uncover meaningful spatial patterns and temporal dynamics in disease outbreaks, providing valuable insights for public health decision-making and resource allocation.
\end{abstract}


\end{frontmatter}

 \section{Introduction}
\label{s:intro}

Disease surveillance relies heavily on accurately interpreting patterns to understand disease stage, manage and prevent health crises.
An important task is to detect neighboring regions with similar disease risk, aiding in identifying populations with high exposure, uncovering inequalities (e.g., in health services, food security), and allocating resources depending on the risk level \citep{lee2012boundary,moraga19}. This becomes even more critical when analyzing disease risk over a period of time; in such cases, a natural extension is to identify neighboring regions with similar functional disease risk within a defined time frame. Identifying such clusters implies detecting neighboring regions where the evolution of a disease or the health status associated with certain conditions is progressing similarly over time, allowing for similar decisions to be made to address the spread or prevalence of the disease of interest and prioritize regions more effectively. While other definitions of clusters exist when working with the space and time domain, we specifically focus on the detection of \textit{spatially contiguous regions exhibiting similar functional risk over time}, referring to these as spatial functional clusters.


Traditional approaches that detect spatial discontinuities in the disease risk surface, often consider constant or time piece-wise  constant risk \citep{knorr2000bayesian,antoniadis2013clustering, anderson2014identifying}. Other approaches focus on detecting a small group of anomalous regions \citep{kulldorff1997spatial, moraga2016detection}. \citet{adin2019two} extended \cite{anderson2014identifying} two-stage approach by incorporating cluster random effects instead of fixed effects only. \cite{wu2022clustering} improved the cluster identification procedure by considering the uncertainty of the number of cluster using Markov chain Monte Carlo (MCMC). However, \cite{wu2022clustering}'s two-staged approach detects spatial clustering from a list of candidates generating from pre-specified cluster algorithms such as K-means and hierarchical clustering. Whether the true cluster lies on the candidates space can not be guaranteed. Furthermore, fixed coefficients of the model are uniform across clusters, and the cluster-specific term follows a Gaussian Markov random field. This limits the flexibility of the model and requires to fit the full model for each iteration. Most mentioned approaches struggle to cope with the complexity and volume of data, particularly when it involves observations recorded continuously during a time interval at distinct locations, known as spatial functional data \citep{romano2015performance}.


Classical functional data clustering techniques seeks to uncover diverse morphological patterns within the continuous functions that underlie discrete measurements or observations \citep{zhang2023review}. Popular methods include projecting functional data into a series of basis functions and then apply conventional clustering algorithms to the coefficient vectors. Typical non-parametric method are based on a pre-distance or dissimilarity, such as hierarchical clustering and K-means clustering \citep{ferreira2009comparison,mammone2018permutation,zambom2019functional}. Model-based clustering algorithms are based on Gaussian mixture or Gaussian mixture process  \citep{ray2006functional,jiang2012clustering,jacques2014model, zeng2019simultaneous}. Although functional data models help capture complexity of data in time domain, the spatial information, very important in disease mapping, is usually ignored. As the increasing demands for analyzing spatial functional data, clustering algorithms considering spatial information are proposed \citep{martinez2020recent}. Based on the way spatial information is used, those
methods can be broadly divided into two categories:  non-parametric clustering based on the temporal dissimilarity measure with a spatial correlation penalty term \citep[see][]{giraldo2012hierarchical}, and a mixture of distributions with Markov random field priors on mixture of parameters \citep{jiang2012clustering}. However, those methods do not guarantee the spatial contiguity of cluster memberships with flexible shape and size. 


More recently, model-based spanning tree partitioning methods have been proposed to perform spatial clustering \citep{teixeira2015generative,teixeira2019bayesian, li2019spatial}. \citet{luo2021bayesian} introduced a Bayesian random spanning tree (RST) partitioning method. Building on this,  \citet{zhang2023bayesian} combined Bayesian random spanning tree with Gaussian distributed spatial functional data. This algorithm is a model-based generative clustering procedure that iteratively samples the clustering structure using reversible jump Markov chain Monte Carlo (RJ-MCMC) to acommodate changes in the parameter space when the cluster structure is modified. Computing the acceptance probability for the proposed cluster structure relies on the conditional marginal distribution given the cluster structure, which is computationally challenging and can hinder convergence for models with more complex terms.
\citet{zhang2023bayesian} addressed this issue using an additional Gibss sampler for the hyper-parameters given the cluster structure and improved mixing using conjugate priors. However, this approach does not hold for more flexible priors and likelihoods, such as the Poisson distribution with random effects, which is widely used in various models for disease risk mapping.


In this paper, we propose a spatial functional cluster model to extend the use of random spanning trees for spatial functional clustering to cases where the response variable comes from a distribution in the exponential family. This model utilizes within-cluster models belonging to a family of latent Gaussian models, which can represent a wide range of processes including seasonality, smooth terms, autoregressive behavior, random effect, and others.
Given the complexity introduced by the lack of conjugacy in our model, the RJ-MCMC algorithm to sample the cluster structure becomes more intricate. Instead, we employ composition sampling and the Metropolis-Hastings algorithm to perform Bayesian inference. To compute the marginal distribution required for our acceptance probability, we use the integrated nested Laplace approximation (INLA). This approach improves mixing and makes inference feasible for spatial functional clustering with non-Gaussian likelihoods.
We demonstrate the adequacy of this method through simulation studies and illustrate its utility in three real-world applications for disease risk mapping.


The remainder of this paper is organized as follows. Section \ref{Method} presents the basic concepts related to random spanning trees and introduces the main spatial functional clustering model proposed in this article.
Section \ref{s:mcmc} introduces the Bayesian inference framework.
We evaluate the performance of our Bayesian inference algorithm in two simulation studies detailed in Section \ref{s:sim}. Section \ref{s:app} analyzes three real-world applications for COVID-19 and dengue in the United States of America (USA) and Brazil. Finally, Section \ref{s:conclusion} provides a discussion of our approach, results, and potential directions for future work.

\section{Methodology}\label{Method}

\subsection{Preliminary concepts}


Let $\mathcal{D} \subset \mathcal{R}^2$ be the space of interest with $n$ spatially contiguous regions. An undirected graph $\mathcal{G} = (V,E)$ can be defined on $\mathcal{D}$, where $V$ is the set of the nodes representing the regions, and $E$ the edges connecting every pair of nodes for regions that share a boundary. A $\mathbf{path}$ from  node $v_1$ to $v_s$ is defined by distinct connected nodes $v_1, \dots, v_{s}$ and the connecting edges $e_i = (v_i, v_{i+1})$ for $i = 1, \dots, s-1$. If a path starts and ends at the same node, it is called a $\textbf{circuit}$. A graph is $\bf{connected}$ if there exists at least one path between any pair of nodes in $V$. A $\textbf{spanning tree}$ $\mathcal{T}$ with respect to a graph $\mathcal{G}$ is a connected sub-graph that contains all nodes of $\mathcal{G}$ with no circuits for any node. Finally, a $\textbf{minimum spanning tree}$ (MST) is a spanning tree derived from an edge-weighted graph with minimum possible total edge weight.


We consider a spatial cluster with respect to the graph $\mathcal{G}$ as any subset of nodes forming a connected sub-graph $\mathcal{G}_c$, for $c = 1, \dots, C \leq n$ \citep{teixeira2019bayesian}. If a set of disjoint subgraphs $M = \{\mathcal{G}_c: c = 1, \dots, C\}$ satisfies $\mathcal{G} = \cup_{c=1}^C \mathcal{G}_c$, then the collection $M$ is known as a partition with $C$ clusters. For a given graph $\mathcal{G}$, any partition $M$ can be derived from at least one minimum spanning tree $\mathcal{T} \subset \mathcal{G}$ by removing a subset of edges of the MST $\mathcal{T}$ \citep{luo2021bayesian}. Hence, the mechanism to sample different partitions of size $C$ for a given graph $\mathcal{G}$ using minimum spanning trees consists on sampling a MST $\mathcal{T}$ and removing $C-1$ edges. A current partition $M$ can also be modified by removing, adding, or removing and adding edges to obtain a new partition $M^*$, making this approach suitable for sampling partitions in Bayesian clustering models.

\subsection{Bayesian spatial functional clustering model} \label{BRST}


A random spanning tree partition model for spatial functional analysis comprises two main components: a spanning tree partition model that defines the clusters and region memberships, and a model for estimating the functional structure within clusters. In this section, we present both the cluster model and the within-cluster model family.


Consider a set of $n$ regions with an associated graph $G$ that characterizes the connections among them. For a given partition $M$ with $C$ clusters derived from a MST $\mathcal{T} \subset G$, we assume the existence of $C$ latent functions $\{h_{c}(t): c \in \{1,2,\dots, C\}\}$ for  $t \in [0,T]$. These functions are not directly observable but are related to the observations through the mean function. Hence, in the within-cluster model, an observed value $y_{it}$ at time $t$ for region $i = 1, \dots, n$, which belongs to cluster $c_i$, is assumed to come from a exponential family distribution with mean $\mu_{c_i,t}$ and additional hyper-parameters $\vect{\theta}_{c_i}$ such as:
\begin{equation}\label{within-cluster-model}
\begin{split}
    & Y_{it} \mid \mu_{it}, \vect{\theta}_{c_i}, \vect M, C, \mathcal{T}  \stackrel{\text{ind}}{\sim} \pi(\cdot \mid \mu_{it}, \vect{\theta}_{c_i}),\\
    & g(\mu_{it}) = \eta_{it} = h_{c_i}(t) + \epsilon_{it},
\end{split}
\end{equation}
where $g(\cdot)$ is a link function connecting the mean $\mu_{it}$ and the linear predictor $\eta_{it}$, and $\epsilon_{it}$ is an optional error term depending on the specific distribution assumed for $\pi(\cdot)$.

We focus on latent Gaussian functions $h_{c}(t)$ represented as:
\begin{equation}\label{latent-function}
    h_{c}(t) = \alpha_{c} + \vect{\beta}_{c}\vect{Z}_{t} + \sum_{k = 1}^{n_f}f^{(k)}_{c}(t), ~~\text{for}~~ c = 1, \dots, C.
\end{equation}
where $\alpha_{c}$ is the cluster-specific intercept, $\vect{\beta}_{c}$ are cluster fixed effects with respect to covariates $\vect{Z}_{t}$ and $f^{(k)}_{c}(\cdot)$ are zero-mean temporal Gaussian random effects such as the joint density function for a fixed set of times $\vect{t}$, given $\vect{\theta}_{c}$, is normally distributed with precision matrix $\vect{Q}_{f^{(k)}}$. Common examples of $f^{(k)}_{c}(\cdot)$ include independent random effects, random walk processes, autoregressive processes, seasonal random effects, and others. Fast inference for this family of models can be performed using integrated nested Laplace approximation (INLA) \citep{rue2009approximate}.

More concisely, if we define  $\vect{x}$ as the collection of all parameters $\{\alpha_{c}\}, \{\vect{\beta}_{c}\}$ and random effects $\{f^{(k)}_{c}(\cdot)\}$, and the $\vect{\theta} = \{\vect{\theta}_c\}$ as the collection of hyper-parameters, then the within-cluster Bayesian model is completed by defining the priors for the Gaussian random vector $\vect{x}$ and hyper-parameters $\vect{\theta}$, given the partition $\vect{M}$, the number of clusters $C$, and the minimum spanning tree $\mathcal{T}$:
\begin{align}
    \vect{x} \mid \vect{\theta}, \vect M, C, \mathcal{T} & \sim \text{MVN}(\vect{0}, \vect{Q}^{-1}),
    & \pi(\vect{\theta} | \vect M, C, \mathcal{T}) & .
\end{align}

The prior for each hyper-parameter in  $\vect{\theta}$ is defined independently, with only requirement being that it is a proper distribution on the support of the hyper-parameter.

In comparison to the model proposed by \citet{zhang2023bayesian}, our approach encompasses a broader family of models with different distributions and temporal components, while \citet{zhang2023bayesian} assume a Gaussian distribution for the response variable and use conjugate priors to derive the posterior distribution conditional on $\{\vect M, C, \mathcal{T}\}$ as well as the marginal distribution, we employ integrated nested Laplace approximation (INLA) to obtain the both the marginal distribution and the posterior distribution as explained in the following section.

The cluster model involves specifying the model for the triple $\vect{\theta}_{\mathcal{T}} = \{\vect M, C, \mathcal{T}\}$ which defines the cluster membership for nodes (regions) in graph $\mathcal{G}$. We use the Bayesian random spanning tree (BRST) model proposed by \citet{luo2021bayesian}, where $\mathcal{T}$ is a minimum spanning tree constructed using weights $\mathbf{w}$, which can be set fixed or follow a uniform distribution $w_{j \ell} \stackrel{\text{i.i.d.}}{\sim} U(0,1)$, and the prior of the number of clusters $C$ is proportional to a geometric distribution:
\begin{equation}\label{brst-prior-1}
    \begin{split}
        \mathcal{T} & =\operatorname{MST}(\mathbf{w}), 
        \quad \pi\left(C = c\right) \propto(1-q)^{c}.
    \end{split}
\end{equation}
The hyper-parameter $q \in[0,1)$  controls the penalty for obtaining a large number of clusters; $q=0$ leads to a discrete uniform prior on $C$, while $q \rightarrow 1$ imposes a large penalty on large values for $C$. Finally, a uniform conditional prior on the partition $\vect M$ is assumed
\begin{equation}\label{brst-prior}
    \begin{split}
\pi\left(\vect M \mid \mathcal{T}, C\right) \propto 1
    \end{split}
\end{equation}
such that, given the MST $\mathcal{T}$ and number of clusters $C$, there are equal probabilities on selecting $C-1$ out of $n-1$ edges of $\mathcal{T}$ to obtain a partition $M$.

\section{Bayesian Inference} \label{s:mcmc}


Bayesian inference for the spatial functional clustering model described in the previous section is achieved by obtaining samples from the joint posterior density of the within-cluster latent field $\vect{x} = \{\{\alpha_c\}, \{\vect{\beta}_c\}, \{f^{(k)}_c(\cdot)\}\}$, within-cluster hyper-parameters $\vect{\theta} = \{\vect{\theta}_c\}$, and cluster parameters $\vect{\theta}_{\mathcal{T}} = \{\vect M, C, \mathcal{T}\}$. Common approaches use reversible jump Markov chain Monte Carlo (RJ-MCMC) due to the unknown number of clusters and the parameter space dimension changing with respect to the number of clusters $C$ \citep[see][]{luo2021bayesian}. In this approach, at each iteration, a modification on the clustering structure is proposed $\vect{\theta}_{\mathcal{T}} \rightarrow \vect{\theta}_{\mathcal{T}}^*$, such as splitting one cluster into two (birth), merging two adjacent clusters (death), splitting a cluster and merging a cluster simultaneously (change), and updating the minimum spanning tree $\mathcal{T}$ and the hyper-parameters $\vect{\theta}$ (hyper). Optionally, an update to the latent field $\vect{x} \rightarrow \vect{x}^*$ can be proposed if required based on the current move. The proposed cluster parameters $\vect{\theta}_{\mathcal{T}}^*$, and within-cluster latent field $\vect{x}^*$ parameters and hyper-parameters $\vect{\theta}^*$ are then accepted with probability $A$, which depends on the within-cluster goodness of fit for the the current and proposed partition (see Algorithm \ref{alg:sampling}).


A critical aspect in RJ-MCMC sampling is to propose new values $\vect{\theta}_{\mathcal{T}}^*$, $\vect{\theta}^*$ and $\vect{x}^*$, and accept them with probability $A$ such as the algorithm efficiently explore the parameter space. Suppose we have completed $(r-1)$ iterations with values $\vect{\theta}_{\mathcal{T}}^{r-1}$, $\vect{\theta}^{r-1}$ and $\vect{x}^{r-1}$. If we propose a move $\vect{\theta}_{\mathcal{T}}^* \sim \pi(\vect{\theta}_{\mathcal{T}})$, hyper-parameters $\vect{\theta}^* \sim \pi(\vect{\theta} \mid \vect{\theta}_{\mathcal{T}}^*)$ and latent field $\vect{x}^* \sim \pi(\vect{x} \mid \vect{\theta}^*, \vect{\theta}_{\mathcal{T}}^*)$, then the acceptance probability is

\begin{equation} \label{A_1}
    A = \frac{\pi(\vect y \mid \vect{\theta}^{*}, \vect{x}^{*}, \vect{\theta}_{\mathcal{T}}^{*})}{\pi(\vect y \mid  \vect{\theta}^{r-1}, \vect{x}^{r-1}, \vect{\theta}_{\mathcal{T}}^{r-1})} \times 
    \frac{\pi( \vect{\theta}_{\mathcal{T}}^{*})}{\pi( \vect{\theta}_{\mathcal{T}}^{r-1})} \times \frac{\pi(\vect{\theta}_{\mathcal{T}}^{r-1} \mid \vect{\theta}_{\mathcal{T}}^{*})}{\pi( \vect{\theta}_{\mathcal{T}}^{*} \mid \vect{\theta}_{\mathcal{T}}^{r-1})}
\end{equation}
Where $\pi(\vect y \mid \vect{\theta}^*, \vect{x}^*, \vect{\theta}_{\mathcal{T}}^*)$
is the likelihood for the full observed data $\vect{y}$ given  $\vect{\theta}^*$, $\vect{\theta}^*$ and $\vect{\theta}_{\mathcal{T}}^*$; $\pi( \vect{\theta}_{\mathcal{T}}^*)$ is the prior density evaluated at the cluster structure $\{\vect{M}^*, C^*, \mathcal{T}^*\}$; and $\pi( \vect{\theta}_{\mathcal{T}}^{*} \mid \vect{\theta}_{\mathcal{T}}^{r-1})$ is the transition probability from $\vect{\theta}_{\mathcal{T}}^{r-1} \rightarrow \vect{\theta}_{\mathcal{T}}^{*}$.
This approach could be applied to the model presented in Section \ref{BRST}; however, the mixing of the chains are increasingly slow with respect to the number of within-cluster model parameters, and it has low acceptance probabilities given that the proposed values come from the prior specification. Alternatively, if we propose a move $\vect{\theta}_{\mathcal{T}}^* \sim \pi(\vect{\theta}_{\mathcal{T}})$ and parameters $\vect{\theta}^* \sim \pi(\vect{\theta} \mid \vect y, \vect{\theta}_{\mathcal{T}}^*)$ and marginalize over $\vect{x}$, then the acceptance probability is
\begin{equation} \label{mhg-decomp}
    A = \frac{\pi(\vect y \mid \vect{\theta}^{*}, \vect{\theta}_{\mathcal{T}}^{*})}{\pi(\vect y \mid  \vect{\theta}^{r-1}, \vect{\theta}_{\mathcal{T}}^{r-1})} \times 
    \frac{\pi(\vect{\theta}^{*} \mid  \vect{\theta}_\mathcal{T}^{*})}{\pi(\vect{\theta}^{r-1} \mid  \vect{\theta}_\mathcal{T}^{r-1})} \times 
    \frac{\pi( \vect{\theta}_{\mathcal{T}}^{*})}{\pi( \vect{\theta}_{\mathcal{T}}^{r-1})} \times \frac{\pi(\vect{\theta}_{\mathcal{T}}^{r-1} \mid \vect{\theta}_{\mathcal{T}}^{*})}{\pi( \vect{\theta}_{\mathcal{T}}^{*} \mid \vect{\theta}_{\mathcal{T}}^{r-1})} \times
    \frac{\pi(\vect{\theta}^{r-1} \mid  \vect y,  \vect{\theta}_\mathcal{T}^{r-1})}{\pi(\vect{\theta}^{*} \mid \vect y, \vect{\theta}_\mathcal{T}^{*})}.
\end{equation}
This approach improves mixing because the hyper-parameters are proposed based on the conditional posteriors, and the latent field is marginalized. Unfortunately, the conditional posterior $\pi(\vect{\theta} \mid \vect y, \vect{\theta}_{\mathcal{T}})$ and the conditional likelihood $\pi(\vect y \mid \vect{\theta}, \vect{\theta}_{\mathcal{T}})$ are not always available, which limits its use. For example, \citet{luo2021bayesian,zhang2023bayesian} use a similar approach for models where the response variable is assumed to be Gaussian and use conjugate priors for the hyper-parameters.

Given that we are interested in perform inference in a wider family of models with good mixing properties, we sample from the posterior $\pi(\vect{\theta}_\mathcal{T}, \vect{\theta}, \vect{x} \mid \vect{y}) = \pi(\vect{\theta}_\mathcal{T} \mid \vect{y}) \pi(\vect{\theta}, \vect{x} \mid\vect{y}, \vect{\theta}_\mathcal{T})$ using compositional sampling. We first aim to obtain samples from the posterior of the cluster model parameters $\pi(\vect{\theta}_\mathcal{T}\mid \vect{y})$ (see Algorithm \ref{alg:sampling}), and later obtain samples from the conditional posterior $\pi(\vect{\theta}, \vect{x} \mid\vect{y}, \vect{\theta}_\mathcal{T})$. Let consider the move $\vect{\theta}_\mathcal{T}^{r-1} \rightarrow \vect{\theta}_\mathcal{T}^*$, then the acceptance probability is
\begin{equation} \label{mhg}
    A = \frac{\pi(\vect y \mid \vect{\theta}_{\mathcal{T}}^{*})}{\pi(\vect y \mid \vect{\theta}_{\mathcal{T}}^{r-1})} \times \frac{\pi( \vect{\theta}_{\mathcal{T}}^{*})}{\pi( \vect{\theta}_{\mathcal{T}}^{r-1})} \times \frac{\pi(\vect{\theta}_{\mathcal{T}}^{r-1} \mid \vect{\theta}_{\mathcal{T}}^{*})}{\pi( \vect{\theta}_{\mathcal{T}}^{*} \mid \vect{\theta}_{\mathcal{T}}^{r-1})}
\end{equation}
where $\pi(\vect y \mid \vect{\theta}_{\mathcal{T}})$ is the within-cluster marginal likelihood given the cluster structure $\vect{\theta}_{\mathcal{T}}$. This term is the most difficult to obtain with respect to the other terms and it is not always analytically tractable for the models of our interest (e.g. Poisson likelihood). Hence, we use the integrated nested Laplace approximation (INLA) proposed in \cite{rue2009approximate} to compute the marginal likelihood conditional on the cluster structure. We show how to compute the other two components, the ratio of cluster structure, $\frac{\pi( \vect{\theta}_{\mathcal{T}}^{*})}{\pi( \vect{\theta}_{\mathcal{T}}^{r-1})}$, and the transition probability $\frac{\pi(\vect{\theta}_{\mathcal{T}}^{r-1} \mid \vect{\theta}_{\mathcal{T}}^{*})}{\pi( \vect{\theta}_{\mathcal{T}}^{*} \mid \vect{\theta}_{\mathcal{T}}^{r-1})}$ in Section 2 of the Supplementary material.


\begin{algorithm}
\caption{Iterative procedure to sample from $\pi(\vect{\theta}_\mathcal{T}\mid \vect{y})$}
\label{alg:sampling}
\begin{algorithmic}[1]
\Procedure{clusteringsampling}{$\mathcal{G}$, $\pi(\vect{\theta}_{\mathcal{T}}), \pi(\vect{\theta}$})\
\State Using the graph $\mathcal{G}$, initialise the minimum spanning tree $\mathcal{T}$ and the partition $\vect{M}$ with $C$ clusters.
\State \parbox[t]{\dimexpr0.96\linewidth-\algorithmicindent}{
Fit the within-cluster models based on current partition $\vect{M}$ and compute the marginal likelihood $\pi(\vect{y} \mid \vect{M}, C, \mathcal{T})$ given the cluster parameters.
}
\For{each iteration}
    \State \parbox[t]{\dimexpr0.96\linewidth-\algorithmicindent}{
    Randomly select one of the following moves: \textbf{birth}, \textbf{death}, \textbf{change}, or \textbf{hyper}, with probabilities $r_{b}, r_{d}, r_{c}$, and $r_{h}$, respectively.
    }
    \State \parbox[t]{\dimexpr0.96\linewidth-\algorithmicindent}{
    Propose new cluster membership $\vect{M}^*$ with $C^*$ clusters depending on the selected move: 
    \begin{itemize}
        \item \textbf{Birth:} Split an existing cluster ($C^* = C + 1$) through the deletion of an edge in the partition $\vect{M}$.
        \item \textbf{Death:} Merge two adjacent clusters ($C^* = C - 1$) by adding an existing edge in $\mathcal{T}$ into the partition $\vect{M}$.
        \item \textbf{Change:} Merge two adjacent clusters by deleting an edge in $\vect{M}$, then split an existing cluster by adding an edge to $\vect{M}$, ensuring the number of clusters remains constant  ($C^* = C$).
        \item \textbf{Hyper:} Obtain a minimum spanning tree $\mathcal{T}^*$ that is compatible with the current partition $\vect{M}$ for given graph $\mathcal{G}$.
    \end{itemize}
    }
    \State \parbox[t]{\dimexpr0.96\linewidth-\algorithmicindent}{
    Fit the within-cluster models based on proposed partition $\vect{M}^*$ with $C^*$ clusters and compute the marginal likelihood for the proposed move $\pi(\vect{y} \mid \vect{M}^*, C^*, \mathcal{T}^*)$.
    }
    \State \parbox[t]{\dimexpr0.96\linewidth-\algorithmicindent}{
    Compute the acceptance probability $A$ for the proposed move. In case of the \textbf{hyper} move, the acceptance probability is $1$ because it does not change the partition $\vect{M}$.
    }
    \State Accept the proposed move $(\vect{M}^*, C^*, \mathcal{T}^*)$ with probability $A$.
\EndFor
\EndProcedure
\end{algorithmic}
\end{algorithm}


Let consider $\vect{y}_c$ the collection of observations for all regions belonging to cluster $c$, and $\vect{x}_c$, the random field associated to cluster $c$. Then, it holds that the conditional posterior of the hyper-parameters
$\pi(\vect{\theta}\mid \vect{y}, \vect{\theta}_{\mathcal{T}}) = \prod_{c=1}^C \pi(\vect{\theta}_c \mid \vect{y}_c, \vect{\theta}_{\mathcal{T}})$ due to $\vect{\theta}_c$ are conditionally independent given the cluster structure $\vect{\theta}_{\mathcal{T}}$. Under the INLA framework the posterior for the hyper-parameters $\vect{\theta}_c$ given the cluster structure $\vect{\theta}_{\mathcal{T}}$ is approximated as $\pi(\vect{\theta}_c\mid \vect{y}_c, \vect{\theta}_{\mathcal{T}}) \propto \pi(\vect{\theta}_c, \vect{x}_c, \vect{y}_c \mid \vect{\theta}_{\mathcal{T}}) / \tilde{\pi}_{\mathrm{G}}(\vect{x}_c \mid \vect{\theta}_c, \vect{y}_c, \vect{\theta}_{\mathcal{T}})|_{\vect{x}_c={\vect{x}_c}_{mode}}$, where $\tilde{\pi}_{\mathrm{G}}(\vect{x}_c \mid \vect{\theta}_c, \vect{y}_c, \vect{\theta}_{\mathcal{T}})$ is a Gaussian approximation to $\pi(\vect{x}_c \mid \vect{\theta}_c, \vect{y}_c, \vect{\theta}_{\mathcal{T}})$ and ${\vect{x}_c}_{mode}$ is the posterior mode of $\vect{x}_c$ for a given value of $\vect{\theta}_c$ and $\vect{\theta}_{\mathcal{T}}$ \citep{rue2009approximate}. As a consequence, a natural approximation for the marginal likelihood is
\begin{equation}\label{eq:marginal}
    \pi(\vect{y}\mid \vect{\theta}_{\mathcal{T}}) =
    \prod_{c=1}^C \pi(\vect{y}_c \mid \vect{\theta}_{\mathcal{T}})
    \approx \prod_{c=1}^C \left.\int \frac{\pi(\vect{\theta}_c, \vect{x}_c, \vect{y}_c \mid \vect{\theta}_{\mathcal{T}})}{\tilde{\pi}_{\mathrm{G}}(\vect{x}_c \mid \vect{\theta}_c, \vect{y}_c, \vect{\theta}_{\mathcal{T}})}\right|_{\vect{x}_c={\vect{x}_c}_{mode}} \vect{d} \vect{\theta}_c.
\end{equation}
The term inside the integral is computed using numerical integration on the space of the hyper-parameters \citep{rue2009approximate}. Notice that the joint density
$\pi(\vect{\theta}_c, \vect{x}_c, \vect{y}_c\mid \vect{\theta}_{\mathcal{T}})=
\pi(\vect{y}_c \mid \vect{x}_c, \vect{\theta}_c, \vect{\theta}_{\mathcal{T}}) \pi(\vect{x}_c \mid \vect{\theta}_c,  \vect{\theta}_{\mathcal{T}}) \pi(\vect{\theta}_c \mid \vect{\theta}_{\mathcal{T}})$ is decomposed as the product of the $c$-cluster likelihood $\pi(\vect{y}_c \mid \vect{x}_c, \vect{\theta}_c, \vect{\theta}_{\mathcal{T}})$ which is a product of densities from the exponential family, the prior of the $c$-cluster latent field $\pi(\vect{x}_c \mid \vect{\theta}_c,  \vect{\theta}_{\mathcal{T}})$ which is zero-mean Gaussian density, and the priors of the hyper-parameters $\pi(\vect{\theta}_c\mid \vect{\theta}_{\mathcal{T}})$. In addition, if we replace the prior for the latent field $\vect{x}_c$, the following holds
\begin{equation}
    \pi(\vect{y}_c \mid \vect{\theta}_{\mathcal{T}}) \approx (2\pi)^{-r/2} {|\vect{Q}|^*}^{1/2}
    \left.\int \frac{\pi(\vect{y}_c \mid \vect{x}_c, \vect{\theta}_c, \vect{\theta}_{\mathcal{T}}) \exp(-\frac{1}{2}\vect{x}_c^T\vect{Q}\vect{x}_c) \pi(\vect{\theta}_c \mid \vect{\theta}_{\mathcal{T}})}{\tilde{\pi}_{\mathrm{G}}(\vect{x}_c \mid \vect{\theta}_c, \vect{y}_c, \vect{\theta}_{\mathcal{T}})}\right|_{\vect{x}_c={\vect{x}_c}_{mode}} \vect{d} \vect{\theta}_c,
\end{equation}
where $r$ is the rank of $\vect{Q}$ and $|\cdot|^*$ denotes the generalized determinant. The term that can be factored out of the integral is sometimes neglected when computing the marginal likelihood. However, in our approach, this term is important for comparing the marginal likelihoods given different cluster structures as required in the acceptance probability in Equation \eqref{mhg}. We use the R-INLA package for our inference and perform the correction when required.

Our approach to sample from the posterior $\pi(\vect{\theta}_{\mathcal{T}} \mid \vect{y})$ is summarised in Algorithm \ref{alg:sampling}, where the acceptance probability is computed according to equations \eqref{mhg} and \eqref{eq:marginal}.
Note that for a particular move (birth, death or change), there is no need to compute the full marginals $\pi(\vect y \mid \vect{\theta}_{\mathcal{T}}^{*})$ and $\pi(\vect y \mid \vect{\theta}_{\mathcal{T}}^{r-1})$ because the ratio
$\pi(\vect y \mid \vect{\theta}_{\mathcal{T}}^{*}) / \pi(\vect y \mid \vect{\theta}_{\mathcal{T}}^{r-1})$ cancels out all the contributions for the clusters that remain the same. Therefore, we only need to compute the marginals for the the clusters that are being modified. Finally, once the samples from $\pi(\vect{\theta}_{\mathcal{T}} \mid \vect{y})$ are obtained, it is straightforward to obtain samples from $\pi(\vect{\theta}, \vect{x} \mid\vect{y}, \vect{\theta}_\mathcal{T}) = \prod_{c=1}^C \pi(\vect{\theta}_c, \vect{x}_c \mid\vect{y}_c, \vect{\theta}_\mathcal{T})$ using INLA or Markov chain Monte Carlo.


\section{Simulation Study} \label{s:sim}

In this section, we evaluate the performance of our spatial functional clustering model and algorithm in two simulation studies. In the first, we test the model's adequacy when the latent function \(h_c(\cdot)\) is either a polynomial or a flexible non-linear shape, using both Gaussian and Poisson data.   We analyze Gaussian data to verify our model under a simple situation, and Poisson data because our goal is to apply spatial functional clustering for disease risk, where Poisson models are widely used. In the second, we examine the model's performance with imbalanced clusters and some neighboring clusters having similar functional shapes. Here, we assess the Poisson model and a Gaussian model applied to Poisson-transformed data to highlight the importance of our approach for non-Gaussian data under three scenarios. For both simulation studies, we use the same set of 100 spatially contiguous regions  $\mathcal{D}$ within a unit square, generated using Voronoi tessellation, and obtain 100 observations over time for each one.

\subsection{Simulation 1: Polynomial and non-linear shape for Gaussian and Poisson data}

\begin{figure}
    \centering
    \includegraphics[width = 1\textwidth]{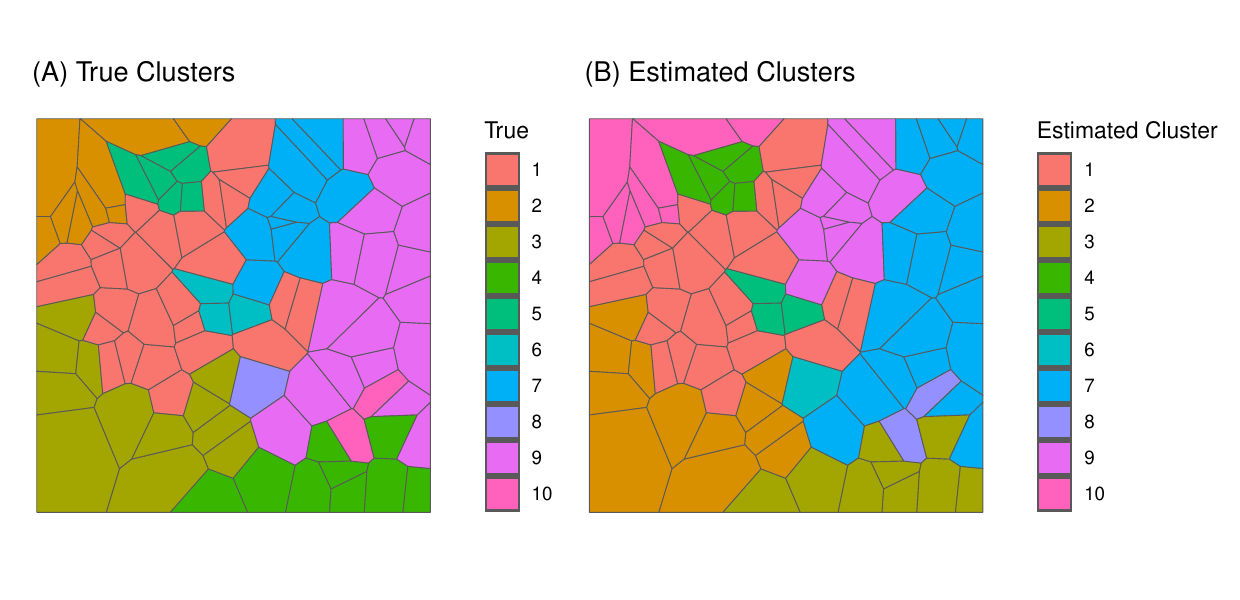}
    \caption{True cluster partition (A) and estimated cluster partition (B) for sub-scenario with flexible non-linear latent function $h_c(t)$ and Poisson data in simulation study 1.}
    \label{fig:rw1_sim1_map}
\end{figure}

In this simulation, we generate $C = 10$  spatial clusters by removing nine edges from a MST generated for the graph $\mathcal{G}$ with respect to $\mathcal{D}$. The generated clusters can be seen in Panel (A) of Figure \ref{fig:rw1_sim1_map}. We create two scenarios: in the first, the latent functions $\{f_c(t): c = 1,\dots, 10\}$ are polynomials, while in the second, they have flexible non-linear shapes.
\begin{itemize}
    \item Polynomial shape: The latent function $h_{c}(t) = \alpha_c + \beta_{c1} t + \beta_{c2} t^2$ is characterized by the cluster parameters $\alpha_c$, $\beta_{c1}$, and $\beta_{c_2}$. The parameters $\beta_c = (\beta_{c1},\beta_{c_2})$ are defined to produce five types of increasing and decreasing functions, and these are repeated between the first five and last five clusters such as $\{\beta_c\}_{c=1}^5 = \{\beta_c\}_{c=6}^{10} = \{(1, 0), (-1, 0), (0,0),(-3, 3), (3, -3)\}$. The parameters $\alpha_c$ are defined so the mean of $h_{c}(t)$, evaluated over the sequence of observed times, is zero.
    \item Flexible non-linear shape: The latent function $ h_{c}(t) = \alpha_c + \sum_{p = 1} ^{16} \vect{B}_p(t) \beta_{cp}$ is defined using 16 basis splines with cluster intercepts $\alpha_c$. The coefficients $\{\beta_{c1}, \dots, \beta_{c,16}\}$ are simulated from an auto-regressive process of order 2 (AR2) with parameters $(0.95, 0)$ for the first five clusters, and $(0.5, 0.44)$ for the remaining clusters. The parameters $\alpha_c$ are defined in a similar way then the polynomial case.
\end{itemize}

For both scenarios, we generate Gaussian and Poisson data, leading to four sub-scenarios. For the Gaussian case, the data for region $i$ at time $t$ is simulated from $Y_{it} \sim \text{N}(h_{c_i}(t), \tau_{c_i}^2)$ with cluster variances $\{\tau_c\} = \{ 0.01, 0.05, 0.02, 0.05, 0.02,0.01, 0.05, 0.02, 0.05, 0.02\}$. For the Poisson case, the data is simulated from $Y_{it} \sim \text{Poisson}(\lambda_{it})$ with mean $\lambda_{it} = \exp(\log(N_i) + h_{c_i}(t) + \epsilon_{it})$ and random effect $\epsilon_{it} \sim N(0,\tau_{c_i}^2)$, where the $\tau_c^2$ values are defined as in the Gaussian case.The population $N_i$ are simulated from a Poisson distribution $Poi(\lambda_i)$, where $log(\lambda_i) \sim N(10, 0.3)$ for all regions.


We executed Algorithm \ref{alg:sampling} with 2000 iterations for Scenario 1 and 3000 iterations for Scenario 2. In both cases, we started with a randomly generated partition having \(c_{0} = 15\) clusters and a hyper-parameter \(q = 0.5\) for the prior distribution of the number of clusters. In Scenario 1 (S1), we fit a within-cluster model where \(h_c(t)\) is represented using fixed effects and monomials. In Scenario 2 (S2), we fit \(h_c(t)\) using a random walk process of order 1.

\begin{figure}
    \centering
        \includegraphics[width = 1\textwidth]{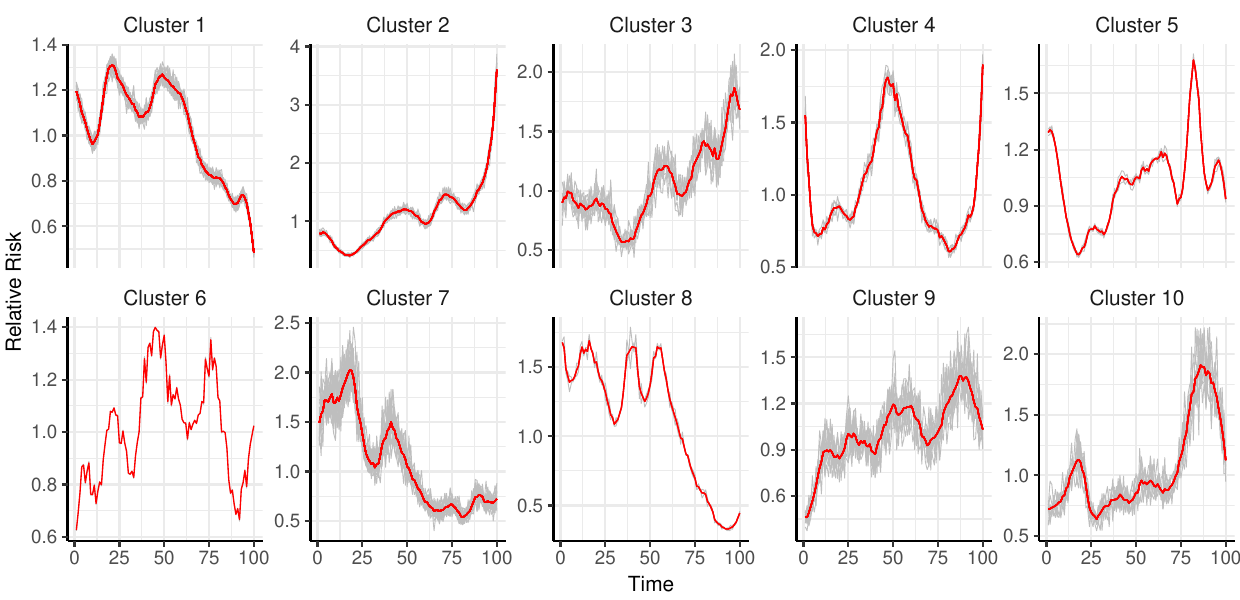}
    \caption{Posterior mean function (red line) for the relative risk, and the observed relative risk (gray lines) of regions per estimated cluster for sub-scenario with flexible non-linear latent function $h_c(t)$ and Poisson data in simulation study 1.}
    \label{fig:rw1_sim2_f}
\end{figure}

For the four sub-scenarios, we successfully recovered the true cluster partition. Once the sampling algorithm reached the true partition, it did not modify it further. Figures \ref{fig:rw1_sim1_map} and \ref{fig:rw1_sim2_f} present the results for the sub-scenario with a non-linear latent function and Poisson data. The results for the other sub-scenarios can be found in Section 3 of the Supplementary Material. Figure \ref{fig:rw1_sim1_map} shows that the estimated partition matches the true cluster partition, with the actual labels of the clusters being irrelevant. Furthermore, Figure \ref{fig:rw1_sim2_f} displays the estimated latent functions alongside the empirical relative risk for each region, demonstrating a common pattern among regions classified within the same cluster.

\subsection{Simulation 2: Imbalanced clusters with similar shape for neighboring clusters}

\begin{figure}
    \centering
    \includegraphics[width = 0.5\textwidth]{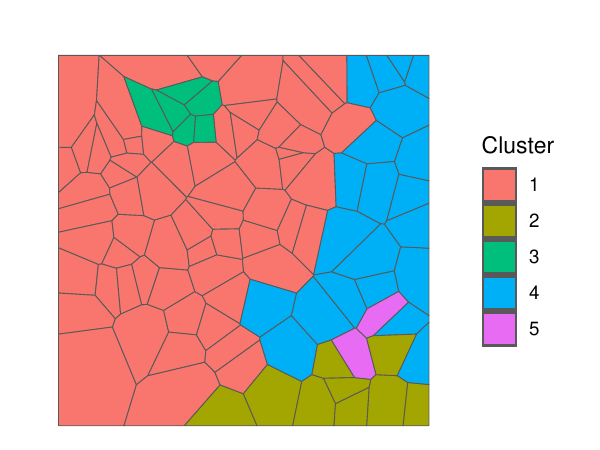}
    \caption{True cluster partition for simulation study 2.}
    \label{fig:sim2-map}
\end{figure}

In this simulation, we partition the regions \(\mathcal{D}\) into \(C = 5\) clusters with an imbalanced number of regions to evaluate our spatial clustering model in a more complex setting (see Figure \ref{fig:sim2-map}). In this partition, Cluster 1 covers \(64\%\) of the regions, Cluster 2 covers around \(10\%\), Cluster 3 around \(6\%\), and Cluster 4 around \(20\%\), while the smallest clusters are Cluster 3 with 6 regions and Cluster 5 with only 2 regions. Notably Cluster 3 is surrounded by Cluster 1, making them prone to merging when the their parameters and hyper-parameters are similar.

\begin{figure}
    \centering
    \includegraphics[width = 1\textwidth]{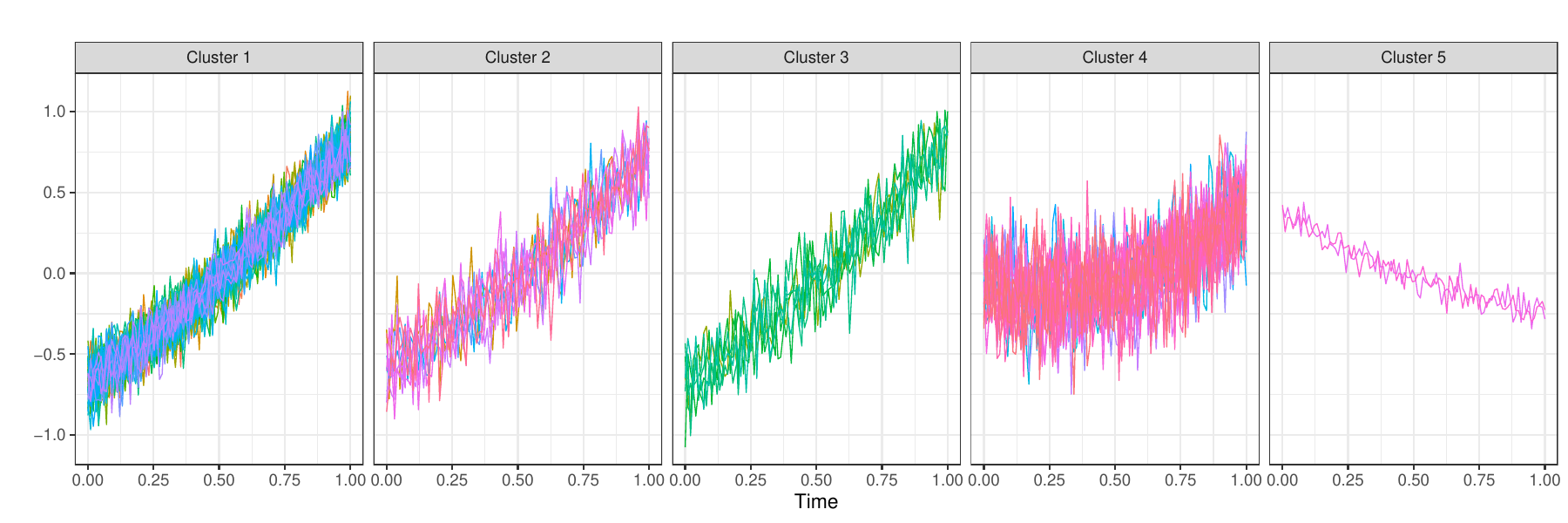}
    \caption{Simulated mean curves per cluster for scenario 2}
    \label{fig:sim2-mean}
\end{figure}

We generate the observation for region \(i\) at time \(t\) from \(Y_{it} \sim \text{Poisson}(\lambda_{it})\) with mean \(\lambda_{it} = \exp(\log(N_i) + \alpha_{c_i} + \beta_{c_i1} t + \beta_{c_i2} t^2 + \epsilon_{it})\) and random effect \(\epsilon_{it} \sim N(0,\tau_{c_i}^2)\). The population \(N_i\) is simulated from a Poisson distribution with mean \(\mu_i\), where \(\log(\mu_i) \sim N(10, 0.3)\) for all regions. The values of the cluster parameters \(\{\beta_{c1}, \beta_{c2}, \tau_{c}\}\) are set to define three scenarios as shown in Table \ref{tab:sim2_param}. In Scenario 1 (S1), the small Cluster 3, surrounded by Cluster 1, has the same trend as Cluster 1 but with higher variability. In Scenario 2 (S2), the trends are the same, but Cluster 3's variability is relatively lower, and the mean trend of Cluster 4 becomes closer to Cluster 1 (see Figure \ref{fig:sim2-mean}). Additionally, the two connected clusters, Cluster 2 and Cluster 5, share the same trend, with Cluster 5 having lower variability. In Scenario 3 (S3), Cluster 1 and Cluster 3 have the same variance but slightly different mean trends. These scenarios are designed to make it more challenging to recover the true clusters.


\begin{table}[h]
    \centering
    \begin{tabular}{ccccccccccc}
    \toprule
    \multirow{2}{*}{Scenario} & \multicolumn{2}{c}{Cluster 1} & \multicolumn{2}{c}{Cluster 2} & \multicolumn{2}{c}{Cluster 3} & \multicolumn{2}{c}{Cluster 4} & \multicolumn{2}{c}{Cluster 5} \\
    \cmidrule(lr){2-3} \cmidrule(lr){4-5} \cmidrule(lr){6-7} \cmidrule(lr){8-9} \cmidrule(lr){10-11}
    & $\beta_c$ & $\tau_c$ & $\beta_c$ & $\tau_c$ & $\beta_c$ & $\tau_c$ & $\beta_c$ & $\tau_c$ & $\beta_c$ & $\tau_c$ \\
    \midrule
    1 & (1, 0.5) & 0.10 & (0.90, 0.4) & 0.10 & (1, 0.5) & 0.15 & (-0.5, 1) & 0.20 & (-1, 0.4) & 0.05 \\
    2 & (1, 0.5) & 0.15 & (-1, 0.4) & 0.10 & (1, 0.5) & 0.12 & (1, 0.8) & 0.20 & (-1, 0.4) & 0.05 \\
    3 & (1, 0.5) & 0.15 & (-1, 0.4) & 0.10 & (0.9, 0.5) & 0.15 & (1, 0.8) & 0.20 & (-1, 0.4) & 0.05 \\
    \bottomrule
    \end{tabular}
    \caption{Cluster parameters for the three scenarios in simulation study 2. The pair \(\beta_c = (\beta_{c1}, \beta_{c2})\) describes the mean structure, and \(\tau_c\) represents the standard deviation.}
    \label{tab:sim2_param}
\end{table}

The performance of the models is evaluated using the Adjusted Rand Index (ARI) and the Normalized Information Distance (NID) suggested by \citet{vinh2009information}. The ARI measures the agreement between two partitions by considering all pairs of elements that are assigned to the same or different clusters \citep{graham1985history}. The ARI is defined as:
\begin{equation}
\begin{split}
   & \text{ARI} = \frac{\text{RI} - \mathbb{E}[\text{RI}]}{\max(\text{RI}) - \mathbb{E}[\text{RI}]}
\end{split}
\end{equation}
where $\text{RI} = (a + b) / \binom{n}{2}$ is the Rand Index, which counts the proportion of pairs that are either in the same cluster (a) or to different clusters $(b)$ in both partitions, out of the total of pairs $\binom{n}{2}$. The ARI ranges from 0, indicating no agreement between the partitions, to 1, indicating perfect agreement.

On the other hand, the NID quantifies the amount of information shared between two partitions relative to the total information contained in the partitions \citep{vinh2009information}. This is defined as:
\begin{equation}
\text{NID} = 1 - \frac{2 I(U,V)}{H(U) + H(V)}
\end{equation}
where $I(U,V) = \sum_{u \in U, v \in V} p(u,v) \log \frac{p(u,v)}{p(u)p(v)}$ is the mutual information between the partitions \(U\) and \(V\), and $H(\cdot) =  -\sum p(\cdot) \log p(\cdot)$ is the entropy. The NID ranges from 0, indicating identical partitions, to 1, indicating dissimilar partitions.

We implemented Algorithm \ref{alg:sampling} with two different likelihoods for all the scenarios: Poisson likelihood with original data and Gaussian likelihood with log-transformed data, keeping the priors of the common parameters the same. We initialized the MST $\mathcal{T}$ from a uniform distributed weights, and generated the partition $M$ by removing 10 edges with top 10 weights, and with a penalty hyper-parameter $q = 0.5$. Then similarly, $h_c(t)$ is expressed as a polynomial function of order 2. For every model, we obtain the final results using Dahl's method after 5000 iterations \citep{dahl2022search}. The method aims to find minimization of
the posterior expected loss using a randomized greedy search algorithm to find a point estimate for a random partition based on a loss function (in our case, we use binder loss) and posterior Monte Carlo samples.

Table \ref{tab:results} presents the outcomes of the three scenarios, while the cluster maps, posterior means by cluster, and the marginal distribution over time are shown in the Supplementary Material. The Poisson model outperforms the Gaussian model with logarithm transformation in all the scenarios, achieving higher ARI and lower NID values. Both models, however, fail to detect the all the true clusters due to the complexity of the scenarios; larger clusters tend to absorb smaller ones when differences between the two are subtle. Nonetheles, the Gaussian model shows a reduced capability to capture imbalanced partition. For instance, Cluster 4 completely disappeared when using the Gaussian model in Scenario 3, whereas the Poisson model successfully identifies it. This indicates that miss-specification of the likelihood may diminish the quality of the clustering.

\begin{table}[htb!]
    \centering
    \caption{Comparison of ARI, RI, NID, and accuracy of the Poisson and the Log-transformed Gaussian model for the three scenarios in simulation study 2.}
    \begin{tabular}{@{} lccccccccc @{}}
        \toprule
        \multirow{2}{*}{Scenario} & \multicolumn{4}{c}{Poisson} & \multicolumn{4}{c}{Normal} \\
        \cmidrule(lr){2-5} \cmidrule(lr){6-9}
        & ARI & RI & NID & Acc & ARI & RI & NID & Acc \\
        \midrule
        Scenario 1 & 0.97 & 0.98 & 0.04 & 99\% & 0.912 & 0.96 & 0.11 & 97\%\\
        Scenario 2 & 0.82 & 0.91 & 0.22 & 94\% & 0.73 & 0.86 & 0.30 & 90\%\\
        Scenario 3 & 0.79 & 0.89 & 0.27 & 92\% & 0.78 & 0.86 & 0.30 & 90\%\\
        \bottomrule
    \end{tabular}
    \label{tab:results}
\end{table}

\section{Application} \label{s:app}

In this section, we apply our proposed spatial clustering framework to three real-world applications. First, we study the weekly incidence of COVID-19 per state in the United States (USA) to identify neighboring states with similar relative risk in 2020. Second, we analyze the weekly incidence of dengue in the state of Minas Gerais, Brazil, between July 2022 and July 2023, aiming to identify neighboring municipalities with similar incidence pattern evolution during this period. Finally, we examine the seasonal behavior of monthly dengue incidence in the state of São Paulo from April 2021 to April 2024. The COVID-19 USA data was obtained from \citet{dong2020interactive}, while the dengue Brazil data was obtained from \citet{CODECO2018S386}.


\subsection{Weekly COVID-19 incidence in the USA in 2020}


In this application, our goal is to identify neighboring states that had similar initial COVID-19 spread in 2020 from a total of 49 states. The COVID-19 pandemic rapidly spread to the United States by early 2020. Detecting spatial-functional data clusters of COVID-19  is crucial for understanding the virus's geographic spread and temporal trends \citep{ribeiroamaraletal23}. By analyzing pre-vaccine COVID-19 data at the state level, we can identify regions with similar infection patterns. This clustering helps public health officials allocate resources more effectively, tailor interventions to specific areas, and anticipate future outbreaks. Such insights are essential for improving preparedness and response strategies, ensuring that measures are both targeted and efficient. 

Considering $Y_{it}$ the number of new cases for state $i$ at time $t$, we use the following within-cluster model:
\begin{equation} \label{brst-uscovid}
\begin{split}
      & Y_{it} \mid \mu_{it}, \vect{\theta}_{\mathcal{T}} ~\stackrel{ind}{\sim}~ \text{Poisson}(E_{i} \times \mu_{it}),\\
     & \log(\mu_{it}) = \eta_{it} = \alpha_{c_i} + f_{c_i,t} + \epsilon_{it}, ~\text{where}~ \epsilon_{it} \mid \vect{\theta}_{\mathcal{T}} {\sim} \mathcal{N}(0, 1/\tau_{c_i}^2),\\
     & f_{c,1} \mid v_{c}, \rho_{c}, \vect{\theta}_{\mathcal{T}} \sim \mathcal{N}(0, v_{c}^{-1}(1-\rho_{c}^2)^{-1}), ~~\text{for}~~ c = 1, \dots, C,\\
     & f_{c, t} = \rho_{c} f_{c, t-1} + \varepsilon_{c,t} ~\text{for}~ t = 2, \dots, T~\text{where}~ \varepsilon_{c,t} \mid \vect{\theta}_{\mathcal{T}} \sim \mathcal{N}(0, v_{c}^{-1}).
\end{split}
\end{equation}
Here, $E_i$ is the expected number of cases for state $i$, and $\{f_{c, t}\}_{t=1}^T$ in an auto-regressive process of order 1 with hyper-parameters $v_{c}$ and $\rho_{c}$. The priors for these parameters are defined as follows: $\log(v_c(1-\rho_c^2)) \mid \vect{\theta}_{\mathcal{T}} \sim \text{LogGamma}(1, 10^{-5})$ and $\log\left(\frac{1+\rho_c}{1-\rho_c}\right) \mid \vect{\theta}_{\mathcal{T}} \sim \mathcal{N}(0, 0.15)$. The priors for the precision hyper-parameters is $\tau_{c}^2 \mid \vect{\theta}_{\mathcal{T}} \sim \text{LogGamma}(1, 5\times 10^{-4})$. 
Finally, the cluster model is as described in Section \ref{BRST} with priors for $\vect{\theta}_{\mathcal{T}} = \{\vect{M}, C, \mathcal{T}\}$ as follows:
\begin{equation} \label{brst-uscovid-priors}
\begin{split}
    &  \mathcal{T}  =\text{MST}(\vect{w}), \quad w_{j \ell} \stackrel{\text { i.i.d. }}{\sim} U(0,1), \quad p\left(C=c\right) \propto 0.5^{c}, \quad p\left( M \mid \mathcal{T}, C\right)  \propto 1.
\end{split}
\end{equation}



We executed the proposed MCMC sampling of Algorithm \ref{alg:sampling} starting with \(c_0 = 10\) clusters, running for 5000 iterations with 3000 iterations as burn-in. The convergence of the algorithm was assessed by analyzing the marginal likelihood (see Figure 25 in the Supplementary Material). Once the chains had converged, we reported the partition selected by the Dahl's algorithm and relabeled the clusters in descending order of size.

\begin{figure}
    \centering
    \includegraphics[width = 0.8\textwidth]{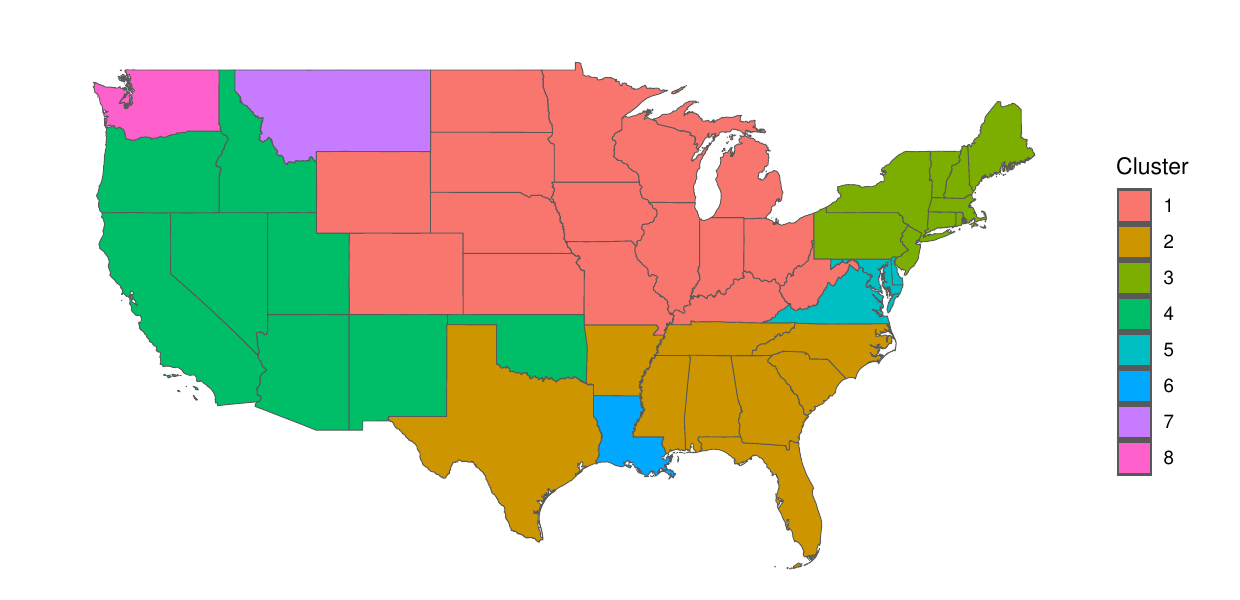}
    \caption{Estimated spatial clusters of COVID-19 relative risk in the USA for 2020.}
    \label{fig:us_covid_map}
\end{figure}

The selected partition after convergence has a total of 8 clusters (Figure \ref{fig:us_covid_map}). The largest clusters are Cluster 1 (C1) to Cluster 4 (C4), located in the north with 16 regions, south with 9 regions, northeast with 9 regions, and west with 8 regions, respectively. Cluster 5 is located in the east with only 4 regions. The remaining clusters (C6-C8) each have only one region (Louisiana, Montana, and Washington, respectively), making them outliers with respect to their neighbors.




\begin{figure}
    \centering
    \includegraphics[width= 1\textwidth]{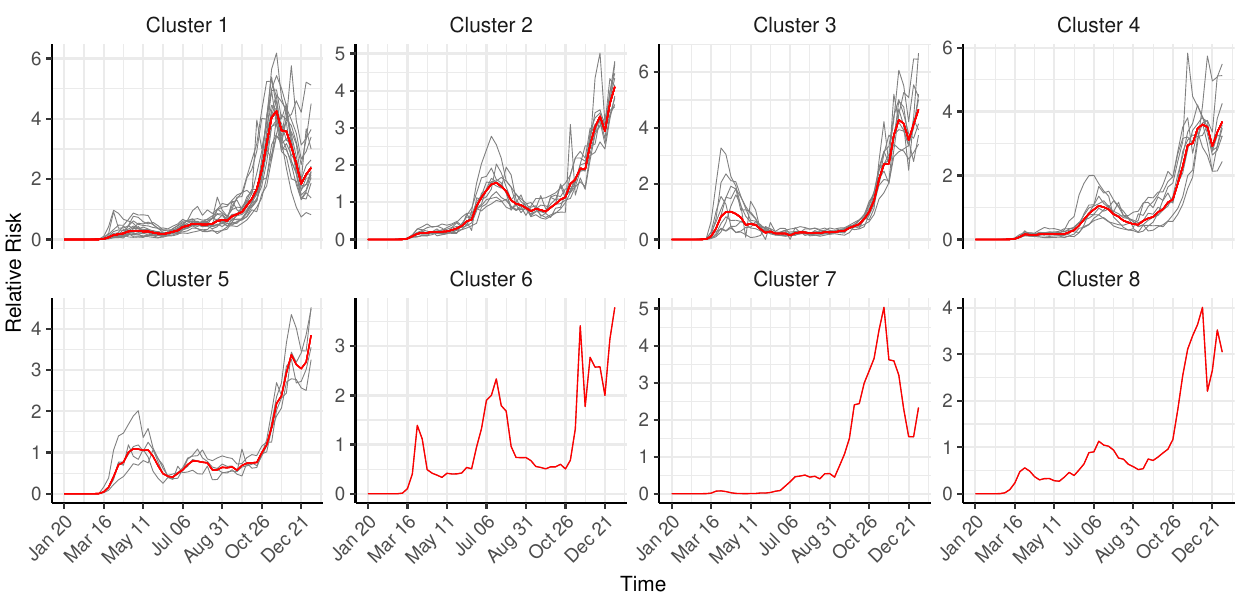} 
    \caption{Predictive mean relative risk (red line) and observed relative risk (gray lines) by estimated spatial cluster for COVID-19 in the USA, 2020.}
    \label{fig:fun_us}
\end{figure}

For all clusters, the relative risk began to increase around March 16 (Figure \ref{fig:fun_us}). Cluster 3 and Cluster 5, neighboring regions in the east zone, experienced an outbreak in April with a higher relative risk than the other clusters. Clusters 2 and 4, located in the south and southwest zones, experienced an outbreak around July 6, with Cluster 2 exhibiting a higher relative risk. Finally, Cluster 1, in the north, experienced a major outbreak in November, while the other clusters began experiencing this outbreak with some delay. These estimated clusters provide insights into the spatial dependencies and variations in virus spread and impact, which is crucial for devising targeted public health strategies tailored to the characteristics of each cluster.

\subsection{Weekly dengue incidence in Minas Gerais, Brazil: July 2022 - July 2023} \label{sub:mg}


Currently, Brazil has experienced a significant rise in dengue cases \citep{WHO2021,koplewitz2022predicting}. The state of Minas Gerais in Brazil tops the country in cases in 2023. Clustering spatial functional data on dengue cases in Minas Gerais is crucial for understanding the spatial-temporal dynamics of the disease, identifying high-risk areas, and optimizing resource allocation for targeted public health responses. To identify neighboring municipalities with similar dengue relative risk evolution, we perform spatial functional clustering in the state of Minas Gerais using weekly dengue cases from $851$ municipalities between July 2022 and July 2023.
 
\begin{figure}
     \centering
     \includegraphics[width  = 1\textwidth]{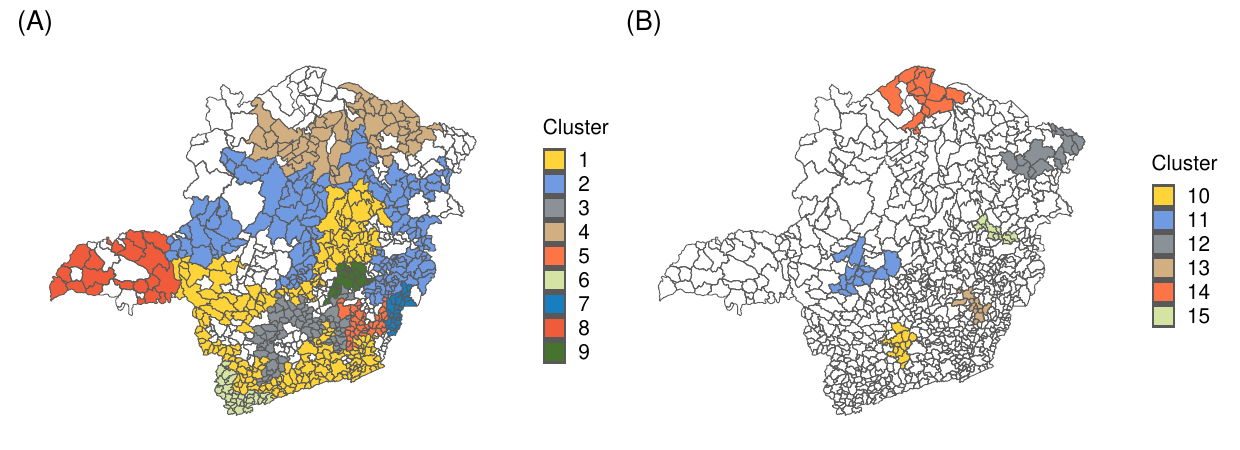}
     \caption{Estimated spatial clusters of dengue relative risk in Minas Gerais, Brazil, between July 2022 and July 2023. It includes only clusters with more than three municipalities.}
     \label{fig:map-mg}
 \end{figure}
\begin{figure}
    \centering
    \includegraphics[width = 1\textwidth]{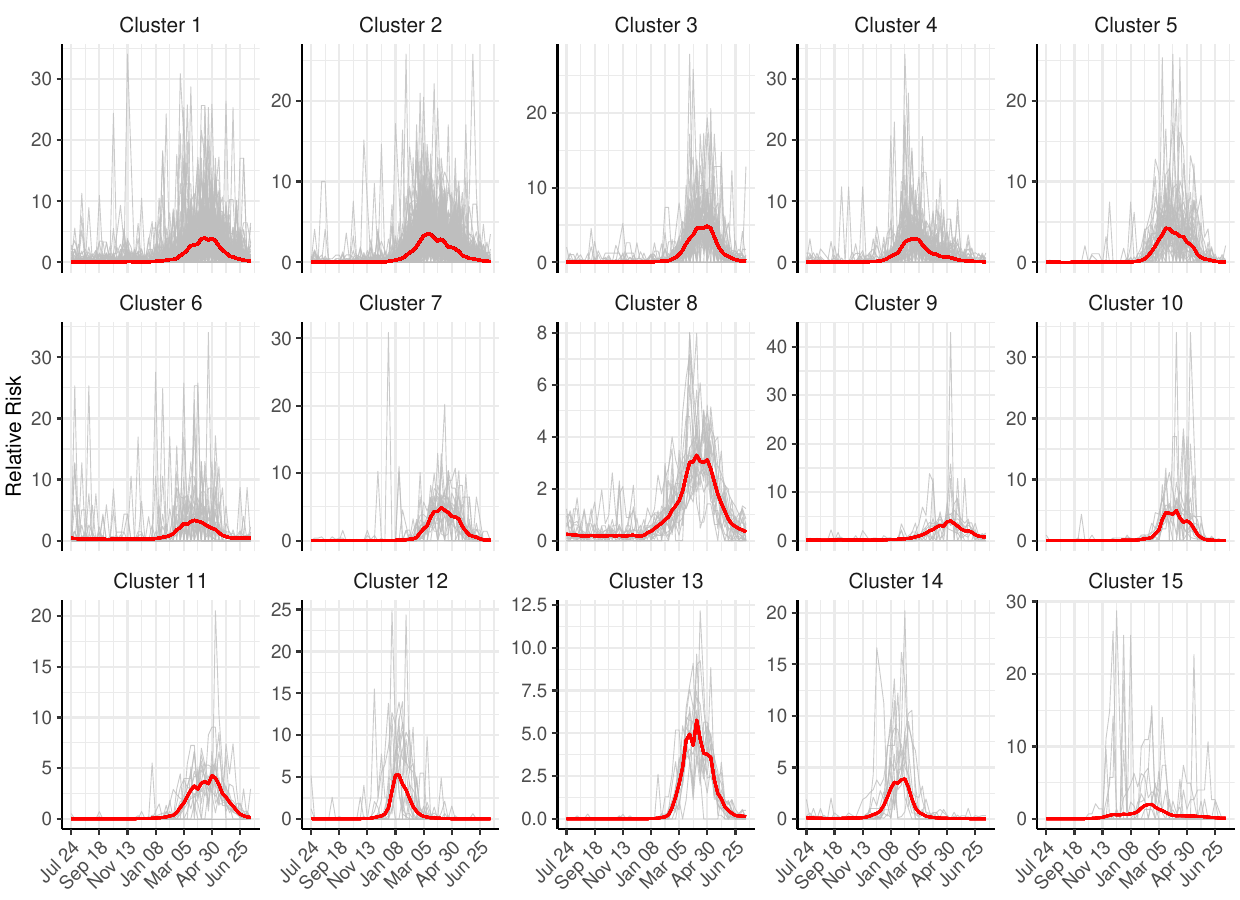}
    \caption{Predictive mean relative risk (red line) and observed relative risk (gray lines) by estimated spatial cluster for dengue in Minas Gerais, July 2022 - July 2023.}
    \label{fig:fun-mg}
\end{figure}

Considering $Y_{it}$ as the number of new cases in region $i$ at week $t$, the model we use is similar to Equation \eqref{brst-uscovid}. However, in this case, the latent effects $\vect{f}_c = (f_{c,1}, f_{c,2}, \dots, f_{c,n})^T$ for cluster $c$ are represented with a random walk process, imposing the conditions $f_{c,t} - f_{c,t-1} \sim \mathcal{N}(0, \nu_{c}^{-1})$ for $t = 2, \dots, n$ such as:
\begin{equation} \label{brst-mg-fprior}
    \pi(\vect{f}_{c} \mid \nu_{c}, \vect{\theta}_{\mathcal{T}}) \propto \nu_c^{(n-1)/2} \exp\left(-\frac{\nu_c}{2}\vect{f}_{c}^T\vect{S}_f\vect{f}_c\right),
\end{equation}
where $\vect{S}_f$ is the structure matrix obtained from the imposed conditions. The prior for the hyper-parameter $\nu_c$ is imposed as follows $\log(\nu_c) \sim \text{LogGamma}(1, 10^{-5})$. The priors for the cluster model is similar to Equation \eqref{brst-uscovid-priors}, differing only in the prior for the number of cluster, $\pi(C= c) \propto 0.999999^{c}$ which imposes higher penalty on a large number of clusters. This means it will favour merge moves over split moves. For this model, Algorithm \ref{alg:sampling} was executed with 100,000 iterations starting with $c_0 = 100$ clusters. The selected partition after convergence is shown in Figure \ref{fig:map-mg}, and the associated latent function for each cluster along with the empirical relative risk is shown in Figure \ref{fig:fun-mg}.


The selected partition after convergence has $63$ clusters. Figure \ref{fig:map-mg} shows the 15 clusters comprising more than three municipalities. Cluster 1 and Cluster 2 are the largest and have very flexible shapes. Cluster 1 is located in the center from east to west, while Cluster 2 covers the center and extends to the west and south. The dynamics of dengue outbreaks between July 2022 to July 2023 for these clusters are shown in Figure \ref{fig:fun-mg}, reveling that 
the dengue outbreak started in the northern part of the state, with relative risk increasing in mid-November, reaching a peak around January 8th, and returning to low levels in March for Clusters 14 and 12. The outbreak in Cluster 4, also in the north, starts in December and lasts until April, while clusters 2 (north) and 8 (west) see an increase in relative risk starting in January. Conversely, the risks in clusters 1, 3, 7, 9, 10, 11, and 13, located between the center and south of Minas Gerais, start to rise after February and return to low levels between mid-May and June.
Overall, it is evident that the outbreak begins in the northern regions and moves southward. However, clusters 15 and 6, despite being located south of clusters 2 and 1, experienced an earlier outbreak, compared to regions with similar locations. 

\subsection{Seasonality of monthly dengue incidence in São Paulo, Brazil: April 2021 - April 2024}

Dengue is a seasonal disease that is strongly influenced by mosquito activity, which depends on environmental and climatic factors \citep{chao2012controlling,pavanietal23}. The seasonal pattern vary across regions due to differing environmental conditions and the presence of micro-regions. Understanding these seasonal variations is essential for making informed public health decisions \citep{monto2010seasonal, 
moraga23}. Therefore, in this application, we aim to cluster $664$ municipalities in the state of São Paulo based on the seasonal patterns observed in the monthly incidence of dengue between April 2021 and April 2024.


Considering $Y_{it}$ as the number of new cases in region $i$ at month $t$, the model we use is similar to Equation \eqref{brst-uscovid}. However, in this case, the expected number of cases $E_{it}$ are computed per year and the latent effects $\vect{f}_c = (f_{c,1}, f_{c,2}, \dots, f_{c,n})^T$ for cluster $c$ are represented with a seasonal random effect, imposing the conditions $f_{c,t} + f_{c,t+1} + \cdots + f_{c,t+m-1} \sim \mathcal{N}(0, \nu_{c}^{-1})$ for $t = 1, \dots, n-m+1$ such as:
\begin{equation} \label{brst-sp-fprior}
    \pi(\vect{f}_{c} \mid \nu_{c}, \vect{\theta}_{\mathcal{T}}) \propto \nu_c^{(n-m+1)/2} \exp\left(-\frac{\nu_c}{2}\vect{f}_{c}^T\vect{S}_f\vect{f}_c\right),
\end{equation}
where $\vect{S}_f$ is the structure matrix obtained from the imposed seasonal conditions. The remaining priors are specified similar to the previous application (Section \ref{sub:mg}). Algorithm \ref{alg:sampling} was executed with $60000$ iterations starting with $c_0 = 200$ clusters.

The selected partition after convergence comprises $79$ clusters. The clusters with more than $5$ municipalities, sorted by size, are shown in Figure \ref{fig:map-sp}, while the remaining clusters are presented in Section 4 of the Supplementary Material. Note that the largest clusters (C1-C6) are located in the center, south, southeast, center-east, northwest, and south of São Paulo, respectively. The estimated relative risk, standardized by year, is shown in Figure \ref{fig:fun-sp}.
Most outbreaks show a seasonal pattern with higher risk between December and June. Cluster 1 and Cluster 5, located in the center and west, respectively, exhibit outbreak peaks around March. Other clusters such as C2-C4 and C6, located in the north, east, and south, exhibit outbreak peaks around April.
The risk in the southeast São Paulo (C3, C4, C6, and C7) reached its peak earlier in 2023 than in 2022, which is not the case for the other clusters.

\begin{figure}
    \centering
    \includegraphics[width = 0.6\textwidth]{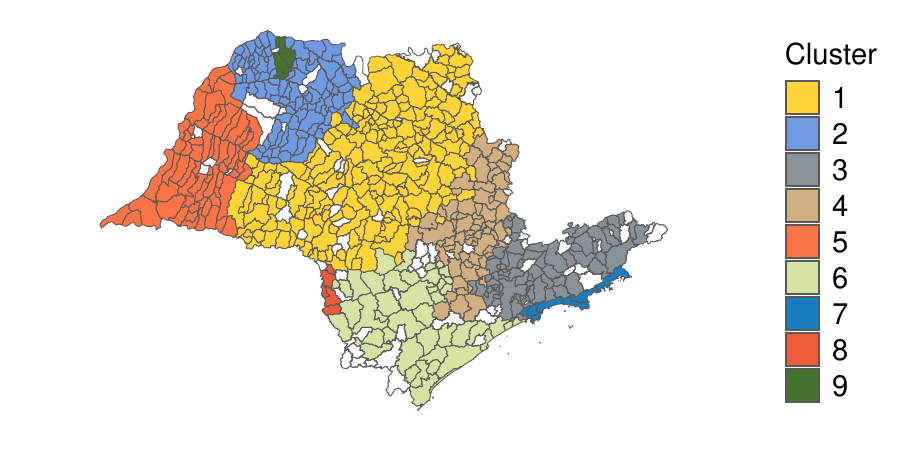}
    \caption{Estimates spatial clusters of dengue seasonal relative risk in São Paulo, Brazil, between April 2021 and April 2024. It includes only clusters with more then five municipalities.}
    \label{fig:map-sp}
\end{figure}

\begin{figure}
    \centering
    \includegraphics[width = 1\textwidth]{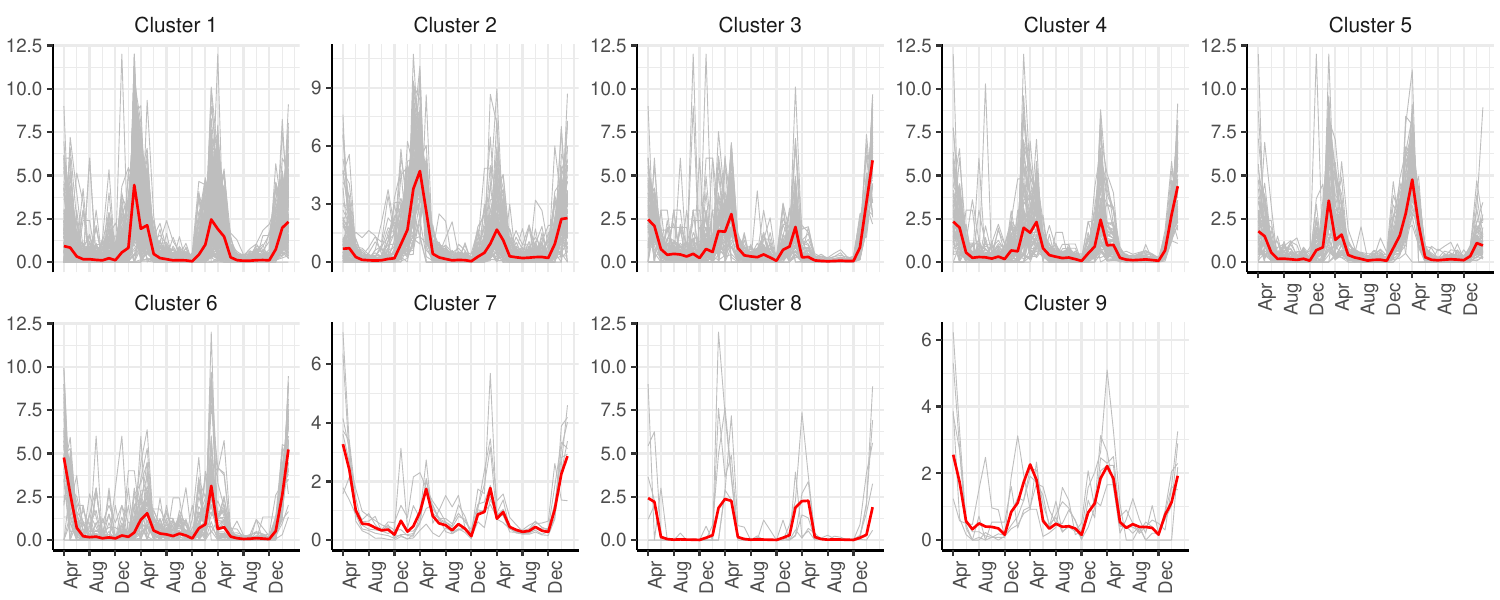}
    \caption{Predicted seasonal mean relative risk (red line) and observed relative risk standardized per year (gray lines) by estimated spatial cluster for dengue in São Paulo, April 2021 - April 2024.}
    \label{fig:fun-sp}
\end{figure}

\section{Conclusion} \label{s:conclusion}


In this article, we propose a spatial functional clustering model for response variables belonging to the exponential family, utilizing random spanning trees for partitioning and latent Gaussian models for the within-cluster structure. Our approach ensures that the resulting clusters comprise neighboring regions with similar latent functions, which can be represented by different processes. We demonstrate the adequacy of our model through simulation studies and then apply it to three real-world applications for disease mapping.
Compared to previous spatial clustering methods in disease mapping, our approach allows all parameters of the within-cluster model to be cluster-specific, resulting in a more flexible setting. Additionally, the number of potential clusters is unlimited, regardless of the cluster's shape and size. Unlike traditional functional clustering algorithms, our method enforces spatial contiguity, constrained by the initial full graph, and can handle non-Gaussian data with latent functions represented by different processes.
This approach enhances our ability to understand, predict, and mitigate outbreak patterns, ultimately reducing morbidity and mortality.

Inference in our spatial functional cluster model is feasible by marginalizing all the parameters associated with the \textit{within-cluster model} and proposing an update in the cluster structure $\vect{\theta}_{\mathcal{T}}$ independent of the within-cluster parameters. In this case, the acceptance probability of our Metropolis-Hasting algorithm depends on the marginal distribution, which is computed using the integrated nested Laplace approximation (INLA), similar to the approach in \citet{gomez2018markov}. Once the partitions are sampled, the within-cluster parameters can be easily sampled conditioned on these partitions.




Our first simulation study shows the algorithm detected the true clusters when the mean functions or hyper-parameters of the clusters differs. In the second simulation, we compared spatial functional cluster models with Poisson likelihood and Gaussian likelihood after a log-transformation. The result shows that the Poisson case performers better than the Normal case suggesting the need of the spatial functional cluster model for different likelihoods. The experiments also illustrates the clustering perform in others challenging situations. Miss-classification could happen when both mean functions and hyper-parameters of different clusters are quite similar, especially when the clusters' sizes are imbalanced. This is reasonable given that the distance between the clusters in parameter space is reduced.


In our first application, we identified clusters of U.S. states with similar relative risk patterns during the COVID-19 outbreak in 2020. Specifically, we found two clusters in the northeast that experienced a major outbreak in April, followed by two clusters that experienced an outbreak in July, and finally, a main cluster in the west that peaked significantly earlier than the others in November. 
In the second application, we observed that dengue outbreaks in Minas Gerais, Brazil, in 2020 began in northern clusters and gradually moved southward. 
Finally, in our last application, we detected municipalities with similar seasonal patterns in São Paulo between 2021 and 2024. In this case, two main clusters, located in the center and west, had peaks around March, while other clusters peaked in April. We also noticed differences in the shape of the seasonal behavior, and in some clusters, the peaks in 2023 occurred earlier than in 2022.


In comparison to classical algorithms, sampling for clusters based on random spanning trees is computationally expensive and might require a large number of iterations when used in complex models with large sample sizes. A significant improvement can be achieved by using adaptive sampling, where the proposal of the cluster structure is modified based on recent samples to enhance the proposed partitions. Our future work will focus on improving the sampling using an adaptive algorithm that employs spanning trees on an auxiliary graph, as recommended by \citet{tam2024exact}.

In conclusion, we proposed a flexible spatial functional clustering method and demonstrated its use in disease surveillance. However, this method is also applicable for clustering in other settings where the response variable comes from a distribution in the exponential family and the within-cluster models belongs to a family of latent Gaussian models.


\pagebreak
\bibliographystyle{imsart-nameyear} 
\bibliography{References}

\end{document}